\documentclass[twocolumn]{aastex62}

\newcommand{\ita}{\textit}
\newcommand{\figs}{Figure~\ref}
\newcommand{\eq}{Equation~\ref}

\newcommand{\quotes}[1]{``#1''}

\newcommand{\bB}{\mathbf{B}}

\graphicspath{{./}{figures/}}

\shorttitle{Flux Ropes and Current Sheets in Solar Wind}
\shortauthors{Pecora et al.}
\begin{document}

\title{Single-spacecraft identification of flux tubes and current sheets in the Solar Wind: combined PVI and Grad-Shafronov method}

\correspondingauthor{Antonella Greco}
\email{antonella.greco@fis.unical.it}

\author{Francesco Pecora}
 \affiliation{Dipartimento di Fisica, Universit\'a  della Calabria, Rende 87036, IT}%Lines break automatically or can be forced with \\
\author{Antonella Greco}%
\affiliation{Dipartimento di Fisica, Universit\'a  della Calabria, Rende 87036, IT}%

\author{Qiang Hu}
\affiliation{University of Alabama Huntsville, Huntsville, AL 35805, USA}

\author{Sergio Servidio}%
\affiliation{Dipartimento di Fisica, Universit\'a  della Calabria, Rende 87036, IT}%

\author{Alexandros G. Chasapis}%
\affiliation{Department of Physics and Astronomy, University of Delaware, Newark, DE, USA}%

\author{William H. Matthaeus}
\affiliation{Department of Physics and Astronomy, University of Delaware, Newark, DE, USA}%



\begin{abstract}

%An approximation method is presented for computing and visualizing the local topology of the magnetic field using single-spacecraft data in the super-Alfv\'enic solar wind.
A novel technique is presented for describing and visualizing the local topology of the magnetic field using single-spacecraft data in the solar wind.
The approach merges two established techniques: the Grad-Shafranov (GS) reconstruction method, which provides a plausible regional two-dimensional magnetic field surrounding the spacecraft trajectory, and the Partial Variance of Increments (PVI) technique that identifies coherent magnetic structures, such as current sheets.
When applied to one month of Wind magnetic field data at 1-minute resolution, we find
that the quasi-two-dimensional turbulence emerges as a sea of magnetic islands and current sheets. Statistical analysis confirms that current sheets associated with high values of PVI are mostly located between and within the GS magnetic islands, corresponding to X-points and internal boundaries. The method shows great promise for visualizing and analyzing single-spacecraft data from missions such as Parker Solar Probe and Solar Orbiter, as well as 1 AU Space Weather monitors such as ACE, Wind and IMAP.

\end{abstract}



\section{Introduction}
The structure of the interplanetary magnetic field at inertial range scales of the observed turbulence is of continuing fundamental and practical interest \citep{GoldsteinEA95,BrunoCarbone05}. Magnetic field turbulence 
influences the propagation of charged particles,
plasma heating, transport of heat, and tangling of magnetic field lines
\citep{MatthaeusVelli11}.
There are also broader fundamental implications for electrodynamics 
in general, and
its applications in astrophysics and plasma laboratory experiments. 
Given the unique opportunity that interplanetary 
spacecraft provide for {\it in situ}
observation, it is important to extract 
as much information as possible from them 
concerning structural properties directly. 
Standard methods 
of time series analysis and 
spectral analysis provide only limited 
information from single-spacecraft 
time series. 
Clusters of satellites provide improved 
information employing 
multi-spacecraft correlation 
techniques \citep{ChhiberEA18}, including ``wave telescope''
\citep{GlassmeierEA01,NaritaEA10-EbAniso} and 
space-time ensemble approaches \citep{MatthaeusEA16}.
Here we present an 
approach for extracting 
additional structural information 
based on a few plausible assumptions, 
and the merger 
of two established techniques: 
the Grad-Shafranov (GS) reconstruction method
\citep{Hau99,Hu2017GSreview} 
which provides 
a plausible regional two-dimensional magnetic field 
topography
surrounding 
the spacecraft trajectory, 
and the Partial Variance of Increments (PVI) technique
\citep{Greco09,greco2018partial} 
that identifies coherent magnetic structures, such as current sheets, as potential magnetic flux tube boundaries.
In this Letter we present a novel combination of these methods, providing new insights into  the nature of magnetic turbulence 
recorded as 
single-spacecraft time series in the Super-Alfv\'enic 
solar wind.

\section{Overview and Background}

Any magnetic field may be partitioned into flux tubes, 
defined generally as cylinders produced by transporting a 
closed contour along the local magnetic field, producing a 
surface everywhere tangent to the field, and well defined 
except at neutral points. Flux ropes are flux tubes carrying a 
current along their magnetic axis. 
Magnetic flux ropes are characterized by their spiral magnetic field-line configurations and have long been studied in heliophysics. 
In studying the nature of magnetic fields in the solar wind, 
a recurrent and central issue is to describe it in terms of the 
flux tubes and flux ropes, field lines being a degenerate case of flux tubes of zero volume.  
Such descriptions account for connectivity as well as constraints 
on the transport of particles, heat and wave energy. 
The so-called ``spaghetti models'' are a 
particular class of observation-based flux tube models
\citep{Schatten71, BrunoEA99-lim, Borovsky08}.
Along boundaries between flux ropes, 
dynamical interactions can produce a 
concentration of gradients,
resulting in structures such as current 
sheets that are approximated as 
directional discontinuities \citep{Greco09}.
Small-scale magnetic flux ropes in the solar wind of durations ranging from a few minutes to a few hours at 1 AU have been identified from in-situ spacecraft data and studied for decades \citep{Moldwin95, Moldwin2000, Feng08, Cartwright10, Yu14}. They possess some similar features in magnetic field configurations to their large-scale counterparts, the magnetic clouds, but, unlike clouds, 
which have a clear solar origin related to coronal mass ejections, the origin of the small-scale magnetic flux ropes is still debated. 
Our view, supported not only by observational analysis but also extensively by numerical simulations over a wide range of scales \citep{Greco09, Servidio11}, 
maintains 
that the presence of 
small-scale magnetic flux ropes or  islands 
is intrinsic to strictly two-dimensional (2D)
magnetohydrodynamic (MHD) turbulence. 
Small-scale flux ropes are believed to be the byproduct of the solar wind turbulence dynamic evolution process, resulting in the generation of coherent structures including \quotes{small random current}, \quotes{current cores}, and \quotes{current sheets} \citep{MattMont80, Veltri99, Greco09} over the inertial range length scales. 
The 2D flux tube paradigm
can be seen to be relevant to the solar wind due to the 
well known fact that plasma turbulence with a large scale mean 
magnetic field tends strongly  
to evolve towards a quasi-2D state \citep{Shebalin83,MattEA90}.
In fact, available observational tests 
repeatedly 
have indicated that solar wind turbulence is 
dominantly quasi-two dimensional
\citep{BieberEA96,HamiltonEA08,MacBrideEA10,NaritaEA10-EbAniso,ChenEA12-3Dstruct} to
a reasonable approximation. 
The next step in our 
reasoning is to recall that 
MHD turbulence exhibits a variety of
relaxation processes 
\citep{Taylor74,MattMont80,StriblingMatt91,ServidioEA08-depress}
that tends to minimize or suppress the strength
of nonlinearities
\citep{KraichnanPanda88,ServidioEA08-depress}.  
This leads to states that have reduced values of 
accelerations, with a preference in nearly 
incompressible MHD for attaining approximately  
force-free, Alfv\'enic and Beltrami states \citep{ServidioEA08-depress}, conditions that are realized to some 
degree in solar wind observations \citep{OsmanEA11-align,ServidioEA12}. 
Local relaxation of these types is fast, less
than a non-linear time, and leads to turbulence 
states that are dynamic but in 
approximate force balance. The evolution of such states can be formally {\it slow}, that is, much slower than the 
time for initial fast relaxation, 
so that quasi-static    
force balance is a reasonable first approximation, 
except at boundaries where coherent structures 
such as current sheets form
\citep{ServidioEA08-depress} 
between relaxed patches.
The above conditions - quasi-two dimensionality, 
and quasi-static force balance - 
justify the GS analysis approach employed below, 
while also providing a relatively clean framework 
for interpretation of the PVI method. 
These methods provide complementary views of
the local structure of the turbulent interplanetary magnetic
field, as we now demonstrate.

\section{GS and PVI Methods}

Reconstruction of 2D, time-independent field and plasma structures from data, taken by a single-spacecraft as it passes through the structures, has been frequently used for the analysis and interpretation of space data \citep{Teh10}. 
The method employed here, 
the Grad-Shafranov (GS) reconstruction technique, 
is based
on the plane GS equation (\ref{eq:GS}), 
developed 
to characterize 
space plasma structures from in situ single-spacecraft measurements \citep{Sonnerup96, Hau99, hu2002reconstruction}

\begin{equation}
    \frac{\partial^2 A}{\partial x^2} + \frac{\partial^2 A}{\partial y^2} = -\mu_0 \frac{d}{dA} \left( p + \frac{B_z^2}{2\mu_0} \right)
    \label{eq:GS}
\end{equation}

where $\textbf{A}=A(x,y)\hat{\textbf{z}}$ is the magnetic vector potential and $\mu_0$ is the vacuum magnetic permeability. The transverse pressure $P_t$ is the sum of the plasma ($p$) and magnetic ($B_z^2/2\mu_0$) pressures, and it is a functions of $A$ only.
The reconstruction technique can be summarized as follows. The isotropic transverse pressure is computed by using the GSE components of magnetic field, plasma bulk flow, plasma number density and isotropized temperatures in the interval of interest.
The structure is supposed to move with a certain velocity. The preferred co-moving frame of reference is the deHoffmann-Teller frame \citep{dHT1950,Gosling2011}, where the electric field vanishes and the magnetic field remains stationary from the Faraday's law. 
Indeed, the optimal velocity of this reference frame is obtained by minimizing the convection electric field. 
To determine the reconstruction frame, minimum variance analysis is performed on the measured magnetic field. The $x$ axis of this frame is along the spacecraft trajectory, where the measurements are known, and it is perpendicular to the symmetry axis $z$. Along the $x$ axis the magnetic vector potential $A(x,0)$ is computed by integrating the magnetic field measurements. The analytic form of the transverse pressure $P_t(A)$ is obtained by fitting the scatter plot $P_t$ \textit{vs} $A(x,0)$ with combinations of polynomials and exponentials. At this point the right-hand side of \eq{eq:GS} is obtained, and the whole equation can be solved. An automated numerical solver has been implemented to quicken the procedure. A detailed description of the steps of this method is given in \cite{hu2002reconstruction,Hu2017GSreview} [see also recent variations in the MMS community \citep{Sonnerup16, Hasegawa19}].
The output of the GS method includes three magnetic field components (the out-of-plane component determined by the force balance) 
and electric current density, given 
over a rectangular domain surrounding the spacecraft path. 
In summary, the GS reconstruction relies on the idea that if a snapshot of the turbulent magnetic field is known to be exactly 2D and quasi-static, 
then it is possible to use cuts through the field in one direction, or maybe several cuts, to reconstruct a reasonable facsimile of the turbulence. 
The method is sensitive to island structures (i.e., O points) but not very sensitive to the large gradients 
that can occur near X points. Indeed, the method mostly produces 
current cores, but not the sharp boundaries. 
As follows, we introduce a method for the identification of these discontinuous boundaries.
The PVI (Partial Variance of Increments) technique is
complementary to the GS method as it 
seeks to identify coherent structures, 
or intense current sheets, that are identified as 
flux tube boundaries or cores \citep{GrecoEA08}, 
and for intense signals,
possible reconnection sites \citep{ServidioEA11-recon,OsmanEA14}. 
In its basic form, PVI is applied to a one-dimensional signal, such as a time series obtained in a high-speed flow, as would be seen by a single-spacecraft in the solar wind, or by a fixed probe in a wind tunnel. PVI is essentially a time series of the magnitude of a vector increment with a selected time lag, normalized by its average over a selected period of time:

\begin{equation}
    \mbox{PVI} = \frac{|\Delta \bB|}{\sqrt{<|\Delta \bB|^2>}},
    \label{eq:PVI}
\end{equation}

where $|\Delta \bB|=|\bB(t+\tau)-\bB(t)|$. It depends on three parameters: its cadence, the time lag $\tau$, and the interval of averaging \citep{greco2018partial}. It is a \quotes{threshold} method and, once a threshold, say $\theta $, has been imposed on the PVI time series, 
a collection, or hierarchy of ``events'' can be identified. It has been shown that the probability distribution of the PVI statistic derived from a non-Gaussian turbulent signal strongly deviates from the Probability Density Function of PVI computed from a Gaussian signal, for values of PVI greater than about 3. As PVI increases to values of 4 or more, the recorded ``events'' are extremely likely to be associated with coherent structures and therefore inconsistent with a signal having random phases. The method is intended to be quite neutral regarding the issue of what mechanism generates the coherent structures it detects. Indeed, the method is sensitive to directional changes, magnitude changes, and any form of sharp gradients in the vector magnetic field {\bf B}.
A comprehensive review of the properties of the 
PVI method provides a broad view of its applications
\citep{greco2018partial}.

\begin{figure}
    \centering
    \includegraphics[width=.5\textwidth]{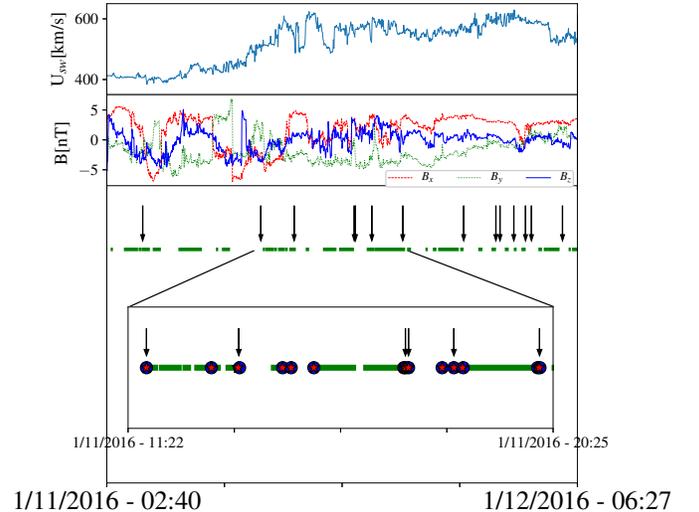}
    \caption{A short time window of the January 2016 dataset ($\sim $ 2 days). The first two top panels display the solar wind bulk speed and magnetic field components. The green horizontal lines represent the duration of the reconstructed flux ropes, the red stars and the blue circles
    (clearly visible in the inset) are the start and end time of PVI events. 
    The black arrows indicate locations of larger PVI values. }
    \label{fig:1d}
\end{figure}
\section{Results}

\begin{figure*}
    \centering
    \includegraphics[width=\textwidth]{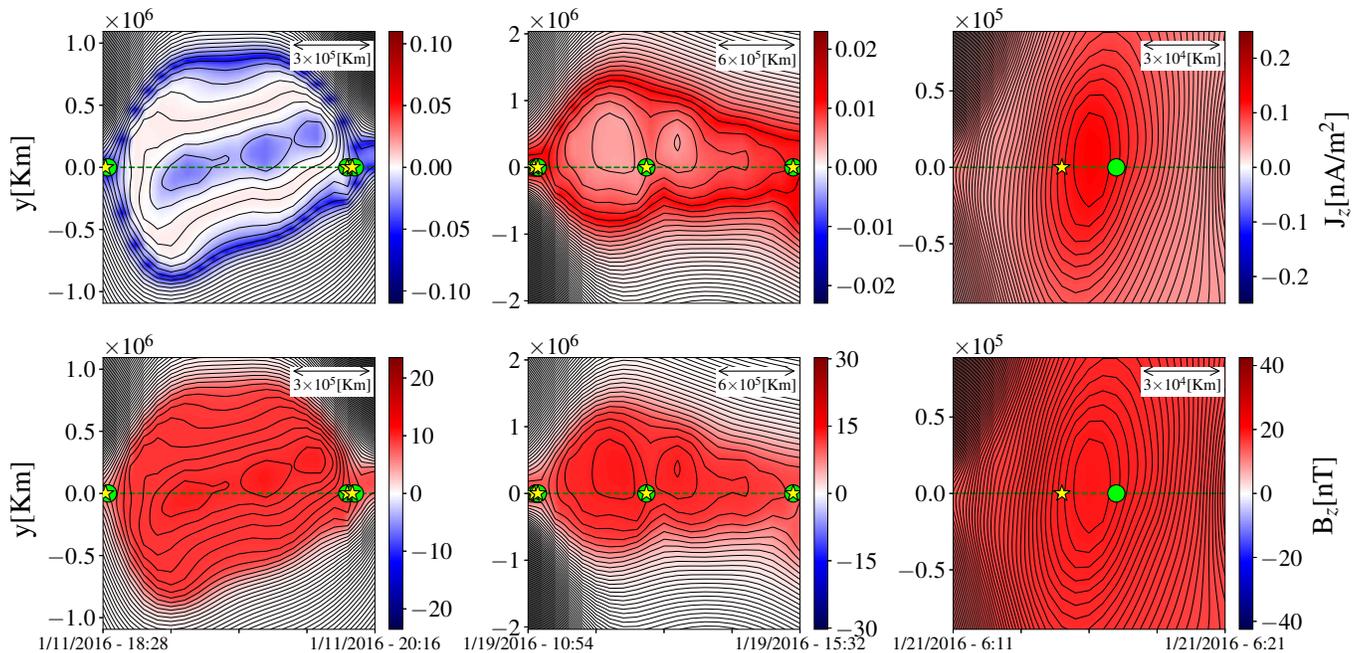}
\vspace{-5mm}
    \caption{Reconstructed flux ropes for Wind January 2016 in the local frame 
   (x,y), with the z-axis representing the cylindrical axis of the flux rope. Magnetic potential contour lines with filled color plots of  $J_z[\mbox{A/m$^2$}]$ (top panels) and $B_z[\mbox{nT}]$ (bottom panels). The dashed lines at $y=0$ are the projection of the spacecraft path on the flux rope cross section. The yellow stars and the green circles represent the start and the end time of the PVI events respectively. 
   The distances in the transverse directions are in km, and they may be considered directly proportional to the magnetic flux across the flux rope. The x-axis of the figures represent the observation period, that we transformed into spatial dimensions (see the ruler inside the plots) applying Taylor hypothesis.}\label{fig:recon}
\end{figure*}

We employ 
{\it in situ} 
measurements of the interplanetary magnetic field and plasma parameters from the Wind spacecraft. Specifically for January 2016, we use the one-minute cadence data sets from the Magnetic Field Investigation (MFI) \citep{Lepping95} and the Solar Wind Experiment (SWE) \citep{Ogilvie95} instruments. All data are accessed via the NASA Coordinated Data Analysis Web (CDAWeb).
For this period, 
we obtained $\sim 400$ magnetic islands or flux tube cross sections via the GS reconstruction and $\sim 400$ PVI events, calculated with a time lag $\tau =2$ min, applying a threshold on the PVI signal $\theta \sim 3.7$. The average that appears in the denominator of \eq{eq:PVI} has been computed over the whole data set \citep{Servidio11}.
In \figs{fig:1d} we show an example of how the PVI and the GS methods work in synergy. The plot is a
1-dimensional view of about 2 day data, indicating the 
occurrence of flux ropes, with PVI events interposed, along with solar wind bulk speed and magnetic field components in the same period. 
The arrows point at the more extreme PVI events that clearly appear at the borders of the magnetic islands. 
The inset provides an expanded view of a 
shorter period of about 9 hours. 
The requirements coming for applicability of 
the GS equations 
in effect select
time intervals during which a possible flux rope appears. 
In appropriate cases, 
we can improve our understanding of the magnetic field topology 
through GS reconstruction, which provides
plausible 2D cross-sections of nearby magnetic islands. 
Meanwhile, the PVI method selects intervals that are suggestive of strong 
currents, which could be cores, but for very strong cases 
tend to be current sheets. 
Applying both methods simultaneously 
enables plausible identification of both flux tubes and 
coherent current structures at their boundaries and it can be done in any single (or multiple)-spacecraft measurement.
Some examples of reconstructed flux ropes in 2D boxes (21$\times$141 points) and magnetic structures 
detected with the PVI method are shown in \figs{fig:recon}, extracted from the same
month of analyzed Wind data. The cross sections of the flux ropes are represented in the local reconstruction frame ($x$, $y$, $z$), with the $z$-axis pointing arbitrarily in space, representing the cylindrical axis of the flux rope.

\begin{figure}
    \centering
    \includegraphics[width=.4\textwidth]{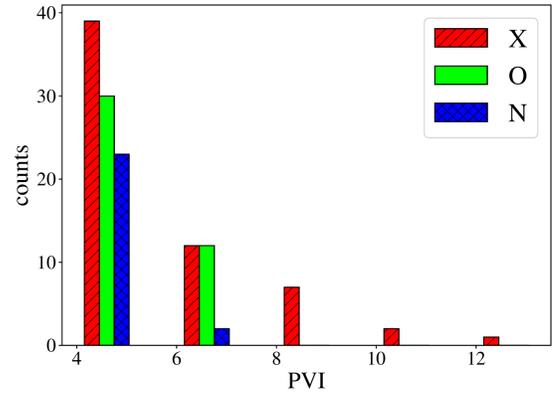}
    \caption{Histograms of PVI events classified as X points (striped red), O points (plain green) and N points (crossed blue). The histogram bins have the same width equal to $2$, starting from PVI$=3.7$.}
    \label{fig:histo}
\end{figure}

Three cases are shown:\\
(I) The leftmost panels display current sheets localized at the 
borders of a large magnetic island, probably X points. 
The lateral extent of this island is of the order of 
2 $\times $10$^6$ km. Its z-axis is mainly in the GSE x-y plane.
(II) In the middle panels 
a longer interval, corresponding to a span of about 
5 nominal correlation scales, is shown. 
Within this very large structure, 
we find two PVI events near the left border of the island and one PVI event within. 
The latter can be interpreted as a core current, however, it is located 
between two secondary islands 
showing the complexity of the magnetic field texture. Here, the local z and GSE z-axes almost coincide.
(III) The rightmost panels show a PVI event found within a  
magnetic island, where the value of $B_z$ is larger.
This is probably a current sheet internal to the flux tube,
associated with bunching of magnetic flux near the 
central axis which is an O point. 
This flux tube is smaller, about  3 $\times $10$^4$ km across, 
or somewhat less than 
the average interplanetary correlation scale. In this case the axis 
of the flux rope points along the GSE y-axis.
In the GS method, the boundaries of a given flux tube are determined by
     the requirement that the pressure-magnetic flux relation $P_t(A)$ remains
     single-valued, and the boundaries appear at points beyond which this can no
     longer be satisfied. In contrast, the PVI method identifies a current sheet
     boundary as a local condition on the vector increment. The boundaries are 
     therefore determined independently  in the two cases, and the finding that they 
     frequently occur in the same or similar positions (see \figs{fig:1d} and \figs{fig:recon}) indicates a synergy in the use of the combined GS/PVI method.
In some reconstructed islands, we were not able to classify the PVI 
events either as X or as O points, and we called them \ita{neither} (N) events. 
One explanation for these
could be that the spacecraft moving in the solar wind may not 
come directly across the X or O points. 
One should be cognizant of the fact that current sheets
in {\it weakly three dimensional}
turbulence \citep{WanEA14}
may also appear within flux tubes but separated both 
from the magnetic axis (core) 
and the X-points that may be found at the boundary with other
flux tubes. We are not aware that such current configurations have been 
reported as emerging in
the purely 2D geometry assumed in the GS method.
Evidently at least an elevation to a weakly 3D reduced MHD model is 
required \citep{RappazzoVelli11,WanEA14}. Nevertheless, 
the GS method may detect signatures of such currents in the solar wind, 
even if this cannot emerge in a purely 2D dynamical model. 
The reconstruction method assumes a local 2D geometry that 
is organized by a strong local, out of plane guide field $B_z$ \citep{OughtonEA94,OughtonEA15},
this state characterized by 
spatial derivatives along the 
$z$ direction that are weak relative to those 
computed in the perpendicular plane. 
Consequently,
a measure of the goodness of the reconstruction
may be given as 
the quantity $\mathcal{A} = \sqrt{\langle\bB_\perp^2\rangle}/{B_z}$ where the averaging operation $\langle\dots\rangle$ is made over a moving 
window. 
By requiring that this quantity is less than $1$ we may
exclude some reconstructed flux ropes and retain more
trustworthy ones. Typical values 
of are of $A$ the order of $0.1$-$0.2$ for the ``good cases''. 
However, in a few cases one finds values of $A$ around $0.6$-$0.8$ 
even if when reconstruction of the flux rope is very good. 
At this point, it is 
interesting to examine 
statistics of the location of the PVI events 
with respect to the magnetic islands. For example, are the more 
intense current sheets occurring at the boundaries of the islands 
(X points)? 
The statistical analysis of over  
$\sim 150$ refined events is shown in \figs{fig:histo}.
The histogram 
confirms 
that the events with the highest PVI values are located at the 
borders of the magnetic islands (where one expects
tangential discontinuities and
possible X points), whereas, 
the cores of the flux ropes (O points) are characterized by a 
broad range of PVI values. The unclassified events are a 
few percentages and at relatively small PVI values.
These may be propagating \citep{WanEA14}
and are perhaps, more likely, rotational discontinuities.
This observational evidence is in agreement with the 
numerical results obtained from a 2D compressible 
MHD simulation shown in \cite{Greco09}.
A physically appealing interpretation emerged: very low values of current lie
     mainly in wide regions (lanes) among magnetic islands. These are associated
     with local low nonlinearity, and possibly wave-like activity \citep{HowesEA18}
     and other transient random currents \citep{Greco16,FranciEA17}. Current cores,
     required by Ampere's law for any flux tube carrying non-zero current, populate
     the central regions of the magnetic islands (or flux tubes). And finally, 
     small-scale current sheet-like structures form narrow regions (sharp boundaries)
     between magnetic islands.
The current sheets represent the well-known small-scale coherent 
structures of MHD turbulence \citep{MattMont80, Veltri99, Servidio08}, 
that are linked to the magnetic field intermittency. This 
classification provides a real-space picture of the nature of 
intermittent MHD turbulence and it found confirmation in the observational data.

We now turn to a comparison   of the electric current densities implied by the GS method 
and the PVI method.  The GS method, within the parameters of its approximations, returns directly, and at each point,
a value 
of the current density.
For the PVI method, we 
can use the empirical result in Fig.~(7) of \citet{greco2018partial} 
as a basis for estimation of the current in the sharp 
boundaries. 
The quoted 
result demonstrates a statistical
relationship 
between 
the normalized 
current density, estimated 
with the curlometer technique \citep{DunlopEA02}, and 
the multi-spacecraft PVI index computed 
from MMS measurements in Earth's magnetosheath. 
The relationship is 
found employing 
normalization of the 
current by $\sigma$, its root mean square value,
a procedure needed to 
compare current measurements with 
PVI, a non-dimensional quantity.
What is suggested in \citet{greco2018partial}
is a strong correlation between PVI 
and current density values that can be expressed as

\begin{figure}
    \centering
    \includegraphics[width=.45\textwidth]{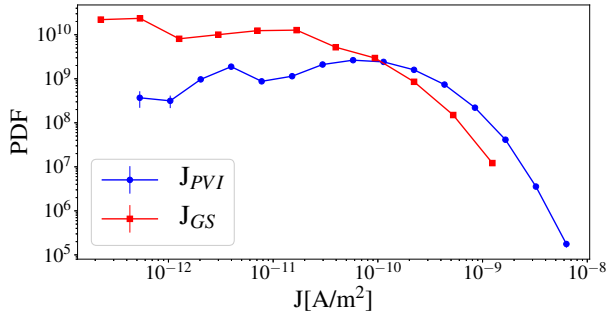}
    \caption{Probability Density Function of the current density evaluated with the GS method (red line) and from PVI signal (blue line).}
    \label{fig:PDFJ}
\end{figure}

\begin{equation}
    \frac{\mbox{J}}{\sigma} \simeq 2 \mbox{ PVI}.
    \label{eq:JPVI}
\end{equation}

Let us suppose that this statistical relationship
applies to the WIND data in the solar wind, 
so we can use \ref{eq:JPVI} to obtain a measure of J. 
To be useful, this procedure requires 
obtaining or estimating the value of 
$\sigma$ in the absence of 
a direct measure of current. 
(otherwise one would not need to use the relation \ref{eq:JPVI}). 
A reasonable estimate of $\sigma$ 
may be obtained 
based on 
computing the root mean square (RMS) value of the 
(single-spacecraft)
measured vector magnetic field increments  
$|\Delta \bf B|$. To convert this 
value to units of current, 
one divides by the magnetic permeability $\mu_0$ and a length L$\sim 10^4$ km, that may be the typical scale of the current sheets. This estimate comes from the statistical distribution of PVI event duration multiplied by the solar wind speed and it is consistent with existing values in literature (e.g. \citet{gosling2008bifurcated}).
In this
approximation $\sigma \mbox{(A/m$^2$)} = \mbox{RMS}(|\Delta {\bf B}|)/(\mbox{L} \mu_0)$, where the average has been computed over the whole data set.
Having $\sigma$ ($\sim 15 \times 10^{-11}$ A/m$^2$) and the entire PVI signal, J values come from the empirical 
expression \ref{eq:JPVI}
\begin{equation}
    \mbox{J}_{PVI}~(\mbox{A/m}^2) \sim 2 \mbox{ PVI } \sigma~(\mbox{A/m}^2).
\end{equation}
The numbers for current density
obtained in this way may be 
compared and contrasted 
with the current obtained, that is implied, 
from the GS reconstruction, say $\mbox{J}_{GS}$,
within each flux rope, sampled along the spacecraft path $(y=0)$.
Accordingly,
we compute the probability density functions 
(PDFs) of $\mbox{J}_{PVI}$ and $\mbox{J}_{GS}$ 
using these two methods 
and illustrate the corresponding 
distributions
in \figs{fig:PDFJ}. 
We emphasize that the currents computed 
from the two
methods are not expected to agree, 
given that the GS current is effectively 
based on the island cores while the 
PVI value is based on the boundaries.
Indeed the figure shows that the $\mbox{J}_{PVI}$ 
distribution is displaced toward 
considerably larger value than the 
$\mbox{J}_{GS}$ distribution.
The most probable GS current occurs at a value that is about two 
orders of magnitude smaller than the most probably PVI current.
The PVI current distribution 
also exhibits a 
noticeable 
extended tail at large values.

\section{Discussion}

The GS reconstruction method and the PVI method provide complementary information when implemented 
together with a single-spacecraft time record of magnetic field.
The results shown here
demonstrate this synergy
with regard to the overall flux tube structure
that is revealed when both methods are employed: 
The GS method is sensitive to the large scale 
magnetic flux tube structure, i.e. the core currents and O-points,
but is not very sensitive to the sharp boundaries, 
the mostly boundary current sheets and X-points. 
The PVI method has the opposite sensitivity, providing 
principally information about localized structures that 
contribute to intermittency, i.e., the flux tube boundaries and associated current sheets.
The two methods identify structure boundaries independently so the result of
     this procedure provides a reasonable, if not purely rigorous, interpretation of the
     local topology of the magnetic field in the region threaded by the observed data.
Another issue is the estimation of electric  
current density. This 
is typically measured in closely spaced
multi-spacecraft missions employing 
a curlometer technique 
\citep{DunlopEA02},
for example in 
the Cluster and Magnetospheric Multiscale (MMS) missions. 
Curlometer is unavailable with single-spacecraft, 
and only with highly sensitive instruments such as 
MMS/FPI (Fast Plasma Investigation instrument) it is possible 
to have direct measurement of current density 
based on the difference in proton/ion 
and electron speeds \citep{PollockEA16}.
However, 
using the present methods one may estimate currents
based on these combined approaches:
GS reconstruction  
provides a 2D picture of the weaker 
flux tube core currents, 
while
the PVI technique along with a 
variance of increments affords 
an estimation of 
the most intense currents at coherent structures. 
We have demonstrated the combined GS/PVI method
and provided additional observational evidence that 
observed solar wind discontinuities 
are coherent structures 
associated with the 
interaction 
of adjacent magnetic flux tubes \citep{GrecoEA08}.  
In particular, this type of 
strong current structure at small scales
is readily interpreted as
a consequence of the intermittent 
nature of fully developed MHD turbulence. \citep{MatthaeusEA15}
The present analysis examines and supports the 
idea that the solar-wind plasma is 
structured \citep{Schatten71,Bruno01, Borovsky08}, 
in the sense that flux tubes are filamentary or 
``spaghetti-like'' (indeed \citet{Hu18} found that the $\hat{z}$ axis of flux ropes at 1AU tend to be aligned with the Parker spiral direction), due to dynamical activity 
in the corona or in interplanetary space. 
With moderate to strong axial fields, 
it is widely acknowledged that 
the plasma tends to become 
locally quasi two dimensional \citep{MattEA90,BieberEA96,HamiltonEA08,MacBrideEA10,
NaritaEA10-EbAniso,ChenEA12-3Dstruct}.
Fast local relaxation \citep{ServidioEA08-depress} also favors the 
2D quasi-static approximation we employed. 
The presence of 
discontinuities or coherent 
current sheets that form between these 
magnetic flux tubes suggests nevertheless 
ongoing dynamical activity, 
and a subset of these might involve magnetic 
reconnection \citep{Dmitruk06,ServidioEA11-recon}. 
These structured flux tubes also provide conduits for 
energetic particle transport and 
possible trapping \citep{TesseinEA13,KhabarovaEA16,PecoraEA18}
and acceleration. 
Indeed one may envision numerous applications in which the combined GS/PVI
     method may reveal structures relevant to understand complex physics and
     turbulent interplanetary dynamics. The method may be particularly useful for
     revealing such features for Parker Solar Probe and Solar Orbiter missions, as
     they explore new regions of the heliosphere in which there are few established
     expectations for the nature of the magnetic field.

\acknowledgments
This research is supported
in part 
by NASA under the MMS Theory and Modeling team (NNX14AC39G),
the Heliophysics Guest Investigator program (NNX17AB79G),
the Parker Solar Probe mission (Princeton 
subcontract SUB0000165). Q. H. work is partially supported by NASA grants 80NSSC18K0623 and 80NSSC19K0276. A. G. and S. S. acknowledges the International Space Science Institute (ISSI) in the framework of International Team 405 entitled “Current Sheets, Turbulence, Structures and Particle Acceleration in the Heliosphere.” 
This work has received funding from the European Unions Horizon 2020 research and innovation programme under grant agreement No 776262 (AIDA, www.aida-space.eu).

\bibliography{biblio}

\begin{thebibliography}{}
\expandafter\ifx\csname natexlab\endcsname\relax\def\natexlab#1{#1}\fi
\providecommand{\url}[1]{\href{#1}{#1}}

\bibitem[{Bieber {et~al.}(1996)Bieber, Wanner, \& Matthaeus}]{BieberEA96}
Bieber, J.~W., Wanner, W., \& Matthaeus, W.~H. 1996, Journal of Geophysical
  Research: Space Physics, 101, 2511

\bibitem[{{Borovsky}(2008)}]{Borovsky08}
{Borovsky}, J.~E. 2008, Journal of Geophysical Research (Space Physics), 113,
  A08110

\bibitem[{{Bruno} {et~al.}(1999){Bruno}, {Bavassano}, {Bianchini},
  {Pietropaolo}, {Villante}, {Carbone}, \& {Veltri}}]{BrunoEA99-lim}
{Bruno}, R., {Bavassano}, B., {Bianchini}, L., {et~al.} 1999, in ESA Special
  Publication, Vol.~9, Magnetic Fields and Solar Processes, ed. A.~{Wilson} \&
  {et al.}, 1147

\bibitem[{{Bruno} \& {Carbone}(2005)}]{BrunoCarbone05}
{Bruno}, R., \& {Carbone}, V. 2005, Living Reviews in Solar Physics, 2, 4

\bibitem[{{Bruno} {et~al.}(2001){Bruno}, {Carbone}, {Veltri}, {Pietropaolo}, \&
  {Bavassano}}]{Bruno01}
{Bruno}, R., {Carbone}, V., {Veltri}, P., {Pietropaolo}, E., \& {Bavassano}, B.
  2001, \planss, 49, 1201

\bibitem[{{Cartwright} \& {Moldwin}(2010)}]{Cartwright10}
{Cartwright}, M.~L., \& {Moldwin}, M.~B. 2010, Journal of Geophysical Research
  (Space Physics), 115, A08102

\bibitem[{{Chen} {et~al.}(2012){Chen}, {Mallet}, {Schekochihin}, {Horbury},
  {Wicks}, \& {Bale}}]{ChenEA12-3Dstruct}
{Chen}, C.~H.~K., {Mallet}, A., {Schekochihin}, A.~A., {et~al.} 2012, \apj,
  758, 120

\bibitem[{{Chhiber} {et~al.}(2018){Chhiber}, {Chasapis}, {Bandyopadhyay},
  {Parashar}, {Matthaeus}, {Maruca}, {Moore}, {Burch}, {Torbert}, {Russell},
  {Le Contel}, {Argall}, {Fischer}, {Mirioni}, {Strangeway}, {Pollock},
  {Giles}, \& {Gershman}}]{ChhiberEA18}
{Chhiber}, R., {Chasapis}, A., {Bandyopadhyay}, R., {et~al.} 2018, Journal of
  Geophysical Research (Space Physics), 123, 9941

\bibitem[{De~Hoffmann \& Teller(1950)}]{dHT1950}
De~Hoffmann, F., \& Teller, E. 1950, Phys. Rev., 80, 692.
\newblock \url{https://link.aps.org/doi/10.1103/PhysRev.80.692}

\bibitem[{{Dmitruk} \& {Matthaeus}(2006)}]{Dmitruk06}
{Dmitruk}, P., \& {Matthaeus}, W.~H. 2006, Physics of Plasmas, 13, 042307

\bibitem[{Dunlop {et~al.}(2002)Dunlop, Balogh, Glassmeier, \&
  Robert}]{DunlopEA02}
Dunlop, M., Balogh, A., Glassmeier, K.-H., \& Robert, P. 2002, Journal of
  Geophysical Research: Space Physics, 107, SMP

\bibitem[{{Feng} {et~al.}(2008){Feng}, {Wu}, {Lin}, {Chao}, {Lee}, \&
  {Lyu}}]{Feng08}
{Feng}, H.~Q., {Wu}, D.~J., {Lin}, C.~C., {et~al.} 2008, Journal of Geophysical
  Research (Space Physics), 113, A12105

\bibitem[{{Franci} {et~al.}(2017){Franci}, {Cerri}, {Califano}, {Landi},
  {Papini}, {Verdini}, {Matteini}, {Jenko}, \& {Hellinger}}]{FranciEA17}
{Franci}, L., {Cerri}, S.~S., {Califano}, F., {et~al.} 2017, \apj, 850, L16

\bibitem[{{Glassmeier} {et~al.}(2001){Glassmeier}, {Motschmann}, {Dunlop},
  {Balogh}, {Acu{\~n}a}, {Carr}, {Musmann}, {Forna{\c{c}}on}, {Schweda},
  {Vogt}, {Georgescu}, \& {Buchert}}]{GlassmeierEA01}
{Glassmeier}, K.~H., {Motschmann}, U., {Dunlop}, M., {et~al.} 2001, Annales
  Geophysicae, 19, 1439

\bibitem[{{Goldstein} {et~al.}(1995){Goldstein}, {Roberts}, \&
  {Matthaeus}}]{GoldsteinEA95}
{Goldstein}, M.~L., {Roberts}, D.~A., \& {Matthaeus}, W.~H. 1995, \araa, 33,
  283

\bibitem[{Gosling \& Szabo(2008)}]{gosling2008bifurcated}
Gosling, J., \& Szabo, A. 2008, Journal of Geophysical Research: Space Physics,
  113

\bibitem[{Gosling {et~al.}(2011)Gosling, Tian, \& Phan}]{Gosling2011}
Gosling, J., Tian, H., \& Phan, T. 2011, The Astrophysical Journal Letters,
  737, L35

\bibitem[{{Greco} {et~al.}(2008){Greco}, {Chuychai}, {Matthaeus}, {Servidio},
  \& {Dmitruk}}]{GrecoEA08}
{Greco}, A., {Chuychai}, P., {Matthaeus}, W.~H., {Servidio}, S., \& {Dmitruk},
  P. 2008, \grl, 35, L19111

\bibitem[{Greco {et~al.}(2018)Greco, Matthaeus, Perri, Osman, Servidio, Wan, \&
  Dmitruk}]{greco2018partial}
Greco, A., Matthaeus, W., Perri, S., {et~al.} 2018, Space Science Reviews, 214,
  1

\bibitem[{{Greco} {et~al.}(2009){Greco}, {Matthaeus}, {Servidio}, {Chuychai},
  \& {Dmitruk}}]{Greco09}
{Greco}, A., {Matthaeus}, W.~H., {Servidio}, S., {Chuychai}, P., \& {Dmitruk},
  P. 2009, The Astrophysical Journall, 691, L111

\bibitem[{{Greco} {et~al.}(2016){Greco}, {Perri}, {Servidio}, {Yordanova}, \&
  {Veltri}}]{Greco16}
{Greco}, A., {Perri}, S., {Servidio}, S., {Yordanova}, E., \& {Veltri}, P.
  2016, \apj, 823, L39

\bibitem[{Hamilton {et~al.}(2008)Hamilton, Smith, Vasquez, \&
  Leamon}]{HamiltonEA08}
Hamilton, K., Smith, C.~W., Vasquez, B.~J., \& Leamon, R.~J. 2008, Journal of
  Geophysical Research: Space Physics, 113

\bibitem[{{Hasegawa} {et~al.}(2019){Hasegawa}, {Denton}, {Nakamura},
  {Genestreti}, {Nakamura}, {Hwang}, {Phan}, {Torbert}, {Burch}, {Giles},
  {Gershman}, {Russell}, {Strangeway}, {Lindqvist}, {Khotyaintsev}, {Ergun},
  {Kitamura}, \& {Saito}}]{Hasegawa19}
{Hasegawa}, H., {Denton}, R.~E., {Nakamura}, R., {et~al.} 2019, Journal of
  Geophysical Research (Space Physics), 124, 122

\bibitem[{{Hau} \& {Sonnerup}(1999)}]{Hau99}
{Hau}, L.-N., \& {Sonnerup}, B.~U.~{\"O}. 1999, \jgr, 104, 6899

\bibitem[{{Howes} {et~al.}(2018){Howes}, {McCubbin}, \& {Klein}}]{HowesEA18}
{Howes}, G.~G., {McCubbin}, A.~J., \& {Klein}, K.~G. 2018, Journal of Plasma
  Physics, 84, 905840105

\bibitem[{{Hu}(2017)}]{Hu2017GSreview}
{Hu}, Q. 2017, Sci.~China Earth Sciences, 60, 1466

\bibitem[{Hu \& Sonnerup(2002)}]{hu2002reconstruction}
Hu, Q., \& Sonnerup, B.~U. 2002, Journal of Geophysical Research: Space
  Physics, 107, SSH

\bibitem[{{Hu} {et~al.}(2018){Hu}, {Zheng}, {Chen}, {le Roux}, \&
  {Zhao}}]{Hu18}
{Hu}, Q., {Zheng}, J., {Chen}, Y., {le Roux}, J., \& {Zhao}, L. 2018, \apjs,
  239, 12

\bibitem[{Khabarova {et~al.}(2016)Khabarova, Zank, Li, Malandraki, le~Roux, \&
  Webb}]{KhabarovaEA16}
Khabarova, O.~V., Zank, G.~P., Li, G., {et~al.} 2016, The Astrophysical
  Journal, 827, 122

\bibitem[{Kraichnan \& Panda(1988)}]{KraichnanPanda88}
Kraichnan, R.~H., \& Panda, R. 1988, The Physics of fluids, 31, 2395

\bibitem[{{Lepping} {et~al.}(1995){Lepping}, {Ac{\~u}na}, {Burlaga}, {Farrell},
  {Slavin}, {Schatten}, {Mariani}, {Ness}, {Neubauer}, {Whang}, {Byrnes},
  {Kennon}, {Panetta}, {Scheifele}, \& {Worley}}]{Lepping95}
{Lepping}, R.~P., {Ac{\~u}na}, M.~H., {Burlaga}, L.~F., {et~al.} 1995, \ssr,
  71, 207

\bibitem[{MacBride {et~al.}(2010)MacBride, Smith, \& Vasquez}]{MacBrideEA10}
MacBride, B.~T., Smith, C.~W., \& Vasquez, B.~J. 2010, Journal of Geophysical
  Research: Space Physics, 115

\bibitem[{Matthaeus {et~al.}(1990)Matthaeus, Goldstein, \& Roberts}]{MattEA90}
Matthaeus, W.~H., Goldstein, M.~L., \& Roberts, D.~A. 1990, Journal of
  Geophysical Research: Space Physics, 95, 20673

\bibitem[{Matthaeus \& Montgomery(1980)}]{MattMont80}
Matthaeus, W.~H., \& Montgomery, D. 1980, Annals of the New York Academy of
  Sciences, 357, 203

\bibitem[{{Matthaeus} {et~al.}(2016){Matthaeus}, {Parashar}, {Wan}, \&
  {Wu}}]{MatthaeusEA16}
{Matthaeus}, W.~H., {Parashar}, T.~N., {Wan}, M., \& {Wu}, P. 2016, \apj, 827,
  L7

\bibitem[{{Matthaeus} \& {Velli}(2011)}]{MatthaeusVelli11}
{Matthaeus}, W.~H., \& {Velli}, M. 2011, \ssr, 160, 145

\bibitem[{Matthaeus {et~al.}(2015)Matthaeus, Wan, Servidio, Greco, Osman,
  Oughton, \& Dmitruk}]{MatthaeusEA15}
Matthaeus, W.~H., Wan, M., Servidio, S., {et~al.} 2015, Philosophical
  Transactions of the Royal Society A: Mathematical, Physical and Engineering
  Sciences, 373, 20140154

\bibitem[{{Moldwin} {et~al.}(2000){Moldwin}, {Ford}, {Lepping}, {Slavin}, \&
  {Szabo}}]{Moldwin2000}
{Moldwin}, M.~B., {Ford}, S., {Lepping}, R., {Slavin}, J., \& {Szabo}, A. 2000,
  \grl, 27, 57

\bibitem[{{Moldwin} {et~al.}(1995){Moldwin}, {Phillips}, {Gosling}, {Scime},
  {McComas}, {Bame}, {Balogh}, \& {Forsyth}}]{Moldwin95}
{Moldwin}, M.~B., {Phillips}, J.~L., {Gosling}, J.~T., {et~al.} 1995, \jgr,
  100, 19903

\bibitem[{{Narita} {et~al.}(2010){Narita}, {Glassmeier}, {Sahraoui}, \&
  {Goldstein}}]{NaritaEA10-EbAniso}
{Narita}, Y., {Glassmeier}, K.~H., {Sahraoui}, F., \& {Goldstein}, M.~L. 2010,
  \prl, 104, 171101

\bibitem[{{Ogilvie} {et~al.}(1995){Ogilvie}, {Chornay}, {Fritzenreiter},
  {Hunsaker}, {Keller}, {Lobell}, {Miller}, {Scudder}, {Sittler}, {Torbert},
  {Bodet}, {Needell}, {Lazarus}, {Steinberg}, {Tappan}, {Mavretic}, \&
  {Gergin}}]{Ogilvie95}
{Ogilvie}, K.~W., {Chornay}, D.~J., {Fritzenreiter}, R.~J., {et~al.} 1995,
  \ssr, 71, 55

\bibitem[{Osman {et~al.}(2014)Osman, Matthaeus, Gosling, Greco, Servidio, Hnat,
  Chapman, \& Phan}]{OsmanEA14}
Osman, K., Matthaeus, W., Gosling, J., {et~al.} 2014, Physical Review Letters,
  112, 215002

\bibitem[{Osman {et~al.}(2011)Osman, Wan, Matthaeus, Breech, \&
  Oughton}]{OsmanEA11-align}
Osman, K.~T., Wan, M., Matthaeus, W.~H., Breech, B., \& Oughton, S. 2011, The
  Astrophysical Journal, 741, 75

\bibitem[{Oughton {et~al.}(2015)Oughton, Matthaeus, Wan, \&
  Osman}]{OughtonEA15}
Oughton, S., Matthaeus, W., Wan, M., \& Osman, K. 2015, Philosophical
  Transactions of the Royal Society A: Mathematical, Physical and Engineering
  Sciences, 373, 20140152

\bibitem[{Oughton {et~al.}(1994)Oughton, Priest, \& Matthaeus}]{OughtonEA94}
Oughton, S., Priest, E.~R., \& Matthaeus, W.~H. 1994, Journal of Fluid
  Mechanics, 280, 95

\bibitem[{{Pecora} {et~al.}(2018){Pecora}, {Servidio}, {Greco}, {Matthaeus},
  {Burgess}, {Haynes}, {Carbone}, \& {Veltri}}]{PecoraEA18}
{Pecora}, F., {Servidio}, S., {Greco}, A., {et~al.} 2018, Journal of Plasma
  Physics, 84, 725840601

\bibitem[{Pollock {et~al.}(2016)Pollock, Moore, Jacques, Burch, Gliese, Saito,
  Omoto, Avanov, Barrie, Coffey, {et~al.}}]{PollockEA16}
Pollock, C., Moore, T., Jacques, A., {et~al.} 2016, Space Science Reviews, 199,
  331

\bibitem[{Rappazzo \& Velli(2011)}]{RappazzoVelli11}
Rappazzo, A., \& Velli, M. 2011, Physical Review E, 83, 065401

\bibitem[{{Schatten}(1971)}]{Schatten71}
{Schatten}, K.~H. 1971, Reviews of Geophysics and Space Physics, 9, 773

\bibitem[{{Servidio} {et~al.}(2011){Servidio}, {Greco}, {Matthaeus}, {Osman},
  \& {Dmitruk}}]{Servidio11}
{Servidio}, S., {Greco}, A., {Matthaeus}, W.~H., {Osman}, K.~T., \& {Dmitruk},
  P. 2011, Journal of Geophysical Research (Space Physics), 116, A09102

\bibitem[{Servidio {et~al.}(2008)Servidio, Matthaeus, \&
  Dmitruk}]{ServidioEA08-depress}
Servidio, S., Matthaeus, W., \& Dmitruk, P. 2008, Physical review letters, 100,
  095005

\bibitem[{{Servidio} {et~al.}(2008){Servidio}, {Primavera}, {Carbone},
  {Noullez}, \& {Rypdal}}]{Servidio08}
{Servidio}, S., {Primavera}, L., {Carbone}, V., {Noullez}, A., \& {Rypdal}, K.
  2008, Physics of Plasmas, 15, 012301

\bibitem[{Servidio {et~al.}(2012)Servidio, Valentini, Califano, \&
  Veltri}]{ServidioEA12}
Servidio, S., Valentini, F., Califano, F., \& Veltri, P. 2012, Physical review
  letters, 108, 045001

\bibitem[{Servidio {et~al.}(2011)Servidio, Dmitruk, Greco, Wan, Donato, Cassak,
  Shay, Carbone, \& Matthaeus}]{ServidioEA11-recon}
Servidio, S., Dmitruk, P., Greco, A., {et~al.} 2011, Nonlinear Processes in
  Geophysics, 18, 675

\bibitem[{{Shebalin} {et~al.}(1983){Shebalin}, {Matthaeus}, \&
  {Montgomery}}]{Shebalin83}
{Shebalin}, J.~V., {Matthaeus}, W.~H., \& {Montgomery}, D. 1983, Journal of
  Plasma Physics, 29, 525

\bibitem[{{Sonnerup} \& {Guo}(1996)}]{Sonnerup96}
{Sonnerup}, B.~U.~{\"O}., \& {Guo}, M. 1996, \grl, 23, 3679

\bibitem[{{Sonnerup} {et~al.}(2016){Sonnerup}, {Hasegawa}, {Denton}, \&
  {Nakamura}}]{Sonnerup16}
{Sonnerup}, B.~U.~{\"O}., {Hasegawa}, H., {Denton}, R.~E., \& {Nakamura},
  T.~K.~M. 2016, Journal of Geophysical Research (Space Physics), 121, 4279

\bibitem[{Stribling \& Matthaeus(1991)}]{StriblingMatt91}
Stribling, T., \& Matthaeus, W.~H. 1991, Physics of Fluids B: Plasma Physics,
  3, 1848

\bibitem[{Taylor(1974)}]{Taylor74}
Taylor, J.~B. 1974, Physical Review Letters, 33, 1139

\bibitem[{{Teh} {et~al.}(2010){Teh}, {Sonnerup}, {Birn}, \& {Denton}}]{Teh10}
{Teh}, W.-L., {Sonnerup}, B.~U.~{\"O}., {Birn}, J., \& {Denton}, R.~E. 2010,
  Annales Geophysicae, 28, 2113

\bibitem[{{Tessein} {et~al.}(2013){Tessein}, {Matthaeus}, {Wan}, {Osman},
  {Ruffolo}, \& {Giacalone}}]{TesseinEA13}
{Tessein}, J.~A., {Matthaeus}, W.~H., {Wan}, M., {et~al.} 2013, The
  Astrophysical Journall, 776, L8

\bibitem[{{Veltri}(1999)}]{Veltri99}
{Veltri}, P. 1999, Plasma Physics and Controlled Fusion, 41, A787

\bibitem[{Wan {et~al.}(2014)Wan, Rappazzo, Matthaeus, Servidio, \&
  Oughton}]{WanEA14}
Wan, M., Rappazzo, A.~F., Matthaeus, W.~H., Servidio, S., \& Oughton, S. 2014,
  The Astrophysical Journal, 797, 63

\bibitem[{{Yu} {et~al.}(2014){Yu}, {Farrugia}, {Lugaz}, {Galvin}, {Leitner},
  {Moestl}, {Nieves-Chinchilla}, {Luhmann}, \& {Wilson}}]{Yu14}
{Yu}, W., {Farrugia}, C.~J., {Lugaz}, N., {et~al.} 2014, AGU Fall Meeting
  Abstracts, SH31A

\end{thebibliography}
\end{document}